\documentclass[11pt]{article}

\usepackage{graphics}   
\usepackage{endnotes}            
\usepackage{amsfonts}            
\usepackage{amssymb}             
\usepackage{amsthm}              
\usepackage{amsmath}            
\usepackage{algorithm}           
\usepackage{algorithmic}
\usepackage{soul}
\sethlcolor{black}
\usepackage{xcolor}
\makeatletter
\usepackage[letterpaper]{geometry}
\usepackage{rotating}            
\usepackage{color}
\usepackage{multibib}

\usepackage{setspace}                
\usepackage{natbib}
\newcommand*{\ULAS}{\ensuremath{{\hat{\Theta}^U}}} 
\newcommand*{\ILAS}{\ensuremath{{\hat{\Theta}^I}}} 
\newcommand*{\epp}{\ensuremath{{e^{{}+{}}}}}        
\newcommand*{\emp}{\ensuremath{{e^{{}-{}}}}}        
\newcommand*{\ecp}{\ensuremath{{e}}}        

\begin{document}

\title{
Mutual Assent or Unilateral Nomination? A Performance Comparison of Intersection and Union Rules for Integrating Self-reports of Social Relationships \thanks{This research was supported in part by ARO award W911NF-14-1-055.}
}
\author{
Francis Lee\thanks{Department of Sociology, University of California, Irvine} and Carter T. Butts\thanks{Departments of Sociology, Statistics, and EECS, and Institute for Mathematical Behavioral Sciences, University of California, Irvine}
}

\date{1/27/18}
\maketitle

\begin{abstract}
Data collection designs for social network studies frequently involve asking both parties to a potential relationship to report on the presence of absence of that relationship, resulting in two measurements per potential tie.  When inferring the underlying network, is it better to estimate the tie as present only when both parties report it as present or do so when either reports it?  Employing several data sets in which network structure can be well-determined from large numbers of informant reports, we examine the performance of these two simple rules.  Our analysis shows better results for mutual assent across all data sets examined.  A theoretical analysis of estimator performance shows that the best rule depends on both underlying error rates and the sparsity of the underlying network, with sparsity driving the superiority of mutual assent in typical social network settings.

\end{abstract}

\theoremstyle{plain}                        
\newtheorem{axiom}{Axiom}
\newtheorem{lemma}{Lemma}
\newtheorem{theorem}{Theorem}
\newtheorem{corollary}{Corollary}

\theoremstyle{definition}                 
\newtheorem{definition}{Definition}
\newtheorem{hypothesis}{Hypothesis}
\newtheorem{conjecture}{Conjecture}
\newtheorem{example}{Example}

\theoremstyle{remark}                    
\newtheorem{remark}{Remark}


Network inference is the problem of inferring an unknown graph from a set of error and/or missingness-prone observations. This problem is of fundamental importance in the study of social networks, where relationships between individuals, organizations, or other entities must typically be inferred from self or proxy reports, archival materials, or other imperfect source of information. Arguably, the most basic and familiar example of the network inference problem arises when attempting to integrate self-reports from subjects, each of whom is asked to identify all others with whom he or she has a particular relationship (or, in the case of a directed relationship, all others to/from whom he or she respectively sends and/or receives ties). Such data has been widely collected \citep[see, e.g.][]{drabek.et.al:bk:1981,reitz:white:1989,bernard.et.al:ara:1984,killworth.bernard:sn:1979,pattison.et.al:jmp:2000}, and poses a basic challenge for the analyst: given two reports on the state of a given relationship, what is to be done when the subjects disagree? \citet{krackhardt:sn:1987} famously formalized two basic strategies for the analysis of such data (leading to respective estimators of the underlying network): regard an edge as present if either party reports it (the \emph{union rule}); or regard an edge as present if and only if both parties report it (the \emph{intersection rule}). While one or another rule has in some cases been argued to be preferred on substantive grounds, there has been little systematic investigation of how the rules perform on empirical data, and in particular on the relative performance of these rules in inferring network structure under realistic conditions. 

This paper seeks to address this gap, employing interpersonal networks whose complete structures can be well-estimated through hierarchical Bayesian models \citep{butts:sn:2003} to assess the accuracy of these simpler (but more widely applicable) rules. Our findings demonstrate that the intersection rule (``mutual assent'') generally outperforms the union rule (``unilateral nomination'') for the networks studied here - a surprising result, given that our informants are not prone to making one particular type of error over another. We resolve this discrepancy by showing that the sparsity of the network is key to the performance of the two rules, with the intersection rule dominating the union rule for networks in which the opportunities for false positives greatly outweigh the opportunities for false negatives.

\section{Background}

Social network analysis is intrinsically and trivially dependent on the ability to accurately measure the structure of social relationships. Despite the rise of network measurement via online social networks, mobile devices, and other sources of observational data, collection of self-reports via sociometric surveys continues to be a popular method for network measurement in a wide range of settings. At least since the seminal studies of Bernard, Killworth, and Sailer-- who provocatively (if hyperbolically) concluded that ``there is no evidence that people know who their network connections are'' \citep{bernard.et.al:ara:1984}--error from informant observations has been known to be a major challenge in network data collection and subsequent inference. Though subsequent studies \citep[e.g.][]{freeman:1987, romney:1982} have tempered the extremity of this conclusion, it is clear that error rates are substantial enough to warrant concern for social network research. With self-report (i.e informants reporting on their own ties)  remaining a popular method of network data collection, there is an ongoing need for simple methods that can maximize the accuracy of networks inferred from this type of information.

Although highly accurate estimates of network structure can be obtained when many measures of each potential tie are available \citep[e.g., from cognitive social structure data - see][]{butts:sn:2003}, simple self-report designs allow only two observations per edge variable (one for each party involved in the potential edge).  The question, then, is how best to integrate these reports to infer the underlying network.  Although many techniques are possible, we here focus on simple, easily used methods of aggregation that (1) estimate the state of an edge variable as being consistent with informants' reports where they agree, and (2) resolve disagreements via a simple uniform rule.  Such strategies lead to estimators\footnote{\citet{krackhardt:sn:1987} does not explicitly treat the LAS as a family of estimators per se, but employs them in a manner consistent with this interpretation.} referred to by \citet{krackhardt:sn:1987} as \emph{locally aggregated structures} (LAS), the union and intersection rules (U-LAS, I-LAS) being the special cases mentioned above.  LAS estimators can be employed for both both undirected relations (when both parties report on the presence/absence of a single undirected edge) and directed relations (when both parties report on their incoming/outgoing ties); indeed, many networks collected in the former manner are erroneously treated as directed, where a LAS or other estimator of an underlying directed relation should be employed.  The present work is thus applicable to any situation in which we obtain edge observations associated with both potential endpoints.

\subsection{Formal Framework} \label{sec_formal}

To formalize the above, our network inference problem may be summarized as follows.  Let $G=(V,E)$ represent an unknown network of interest, with fixed and known vertex set $V$ and unknown edge set $E$.  Without loss of generality, we will represent $G$ via its adjacency matrix, $\Theta$; where $G$ is undirected, $\Theta$ is constrained to be symmetric.  Here, we assume the vertex set to be fixed and the edge set unknown.  Informant reports are represented via an informant by sender by receiver adjacency array, $Y$, such that $Y_{ijk}=1$ if $i$ reports that the edge from $j$ to $k$ is present (with 0 otherwise).  In our setting, we assume that informants report only on their own ties, and hence only $Y_{iij}$ and $Y_{jij}$ (and their reciprocating edge variables) are employed.  The locally aggregated structure (LAS) introduced by \citet{krackhardt:sn:1987} has been a popular network inference tool for aggregating an ego's and alter's judgments from such data. While this has traditionally been explored through cognitive social structures (which collect an informant's perception of the entire social structure), the only responses \emph{needed} from an informant are the ties they report sending out and the ties they perceive others send to them. In terms of the above, the union and intersection LAS estimators are defined as follows:
\begin{gather}
\ULAS_{ij} = 1-(1-Y_{iij})(1-Y_{jij}) \label{e_ulas} \\
\ILAS_{ij} = Y_{iij} Y_{jij} \label{e_ilas}
\end{gather}
\noindent As noted above, \ULAS estimates an edge as being present when either party reports it, while \ILAS does so only when both parties agree.  Both rules are simple and easily understood, but may lead to very different estimates of network structure.  If one must employ either \ULAS or \ILAS, which should one use?  To determine this, we consider the accuracy of each estimator under realistic conditions.

\section{LAS Accuracy: Some Basic Theory} \label{sec_theory}

\begin{figure}
\begin{center}
\includegraphics[width=2.5in]{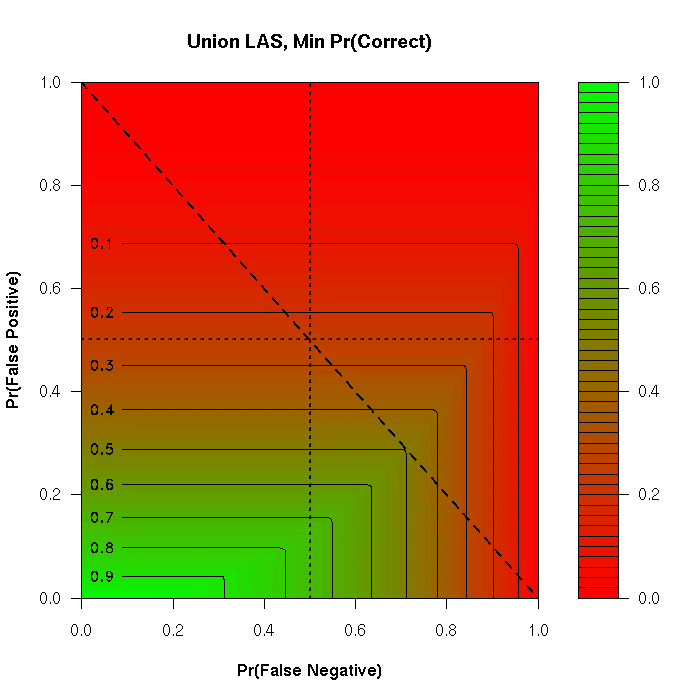}\includegraphics[width=2.5in]{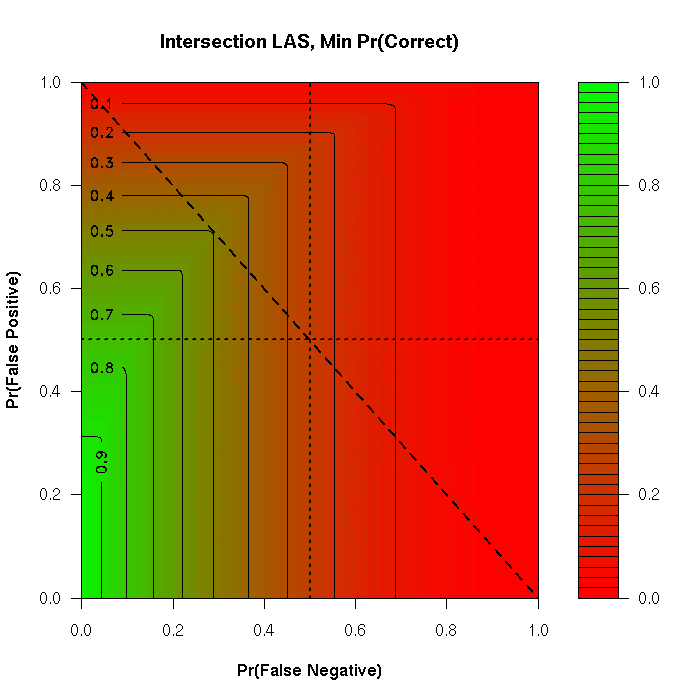}
\caption{Single edge-variable error rates for union and intersection rules, as a function of informant error rates. The intersection rule is robust to false positive errors but vulnerable to false negative errors, with the reverse holding for the union rule. \label{f_base_las_err}}
\end{center}
\end{figure}

It is not immediately clear which LAS method would provide a more accurate estimate of the unknown graph. If informants uniformly make more false positive errors (reporting that an edge exists when it does not) relative to their false negative rate (reporting that an edge does not exist when it does) in their edge observations, then it would seem the LAS intersection would obtain an estimated graph close to the unknown graph. Conversely, if informants make a greater number of false negative errors relative to the false positive rate, then the LAS union would be expected to provide a closer estimate to the unknown graph. This intuition follows from the response of the respective rules to informant error rates on a per-edge basis.  Define the false positive and false negative error rates for an arbitrary informant $i$ by 
\begin{gather*}
\epp_i=\Pr(Y_{iij}=1|\Theta_{ij}=0)=\Pr(Y_{iji}=1|\Theta_{ji}=0)\\
\emp_i=\Pr(Y_{iij}=0|\Theta_{ij}=1)=\Pr(Y_{iji}=0|\Theta_{ji}=1),
\end{gather*}
with $\ecp=(\epp,\emp)$ being the full set of error rates.  Assuming that errors occur independently, it then immediately follows that the per-edge error rates for \ILAS and \ULAS are given by
\begin{gather*}
\Pr(\ILAS_{ij}\neq\Theta_{ij}|\ecp) = \begin{cases} \epp_i\epp_j & \Theta_{ij}=0\\ 1-(1-\emp_i)(1-\emp_j)& \Theta_{ij}=1\end{cases} \\
\Pr(\ULAS_{ij}\neq\Theta_{ij}|\ecp) = \begin{cases}  1-(1-\epp_i)(1-\epp_j) & \Theta_{ij}=0\\ \emp_i\emp_j& \Theta_{ij}=1\end{cases}. 
\end{gather*}
Where error rates are approximately equal across informants, \ILAS{} false positive rates scale with the square of the individual false positive rates, with the same holding mutatis mutandis for \ULAS and false negative rates; while this can result in substantial suppression for these types of errors, errors of the opposite type (false negatives for \ILAS, false positives for \ULAS) are correspondingly magnified.  This is illustrated graphically in Figure~\ref{f_base_las_err}, which shows the ``worst case'' probabilities of a correct inference as a function of informant accuracy (with both informants assumed to have the same error rates).  In the typical setting for which $\emp>\epp$---i.e., omission of true ties, due e.g. to forgetting, is more common than fabrication or confabulation of nonexistent ties----this analysis suggests that $\ULAS$ should be more accurate than $\ILAS$ (perhaps by a large margin).

\begin{figure}
\begin{center}
\includegraphics[width=4.5in]{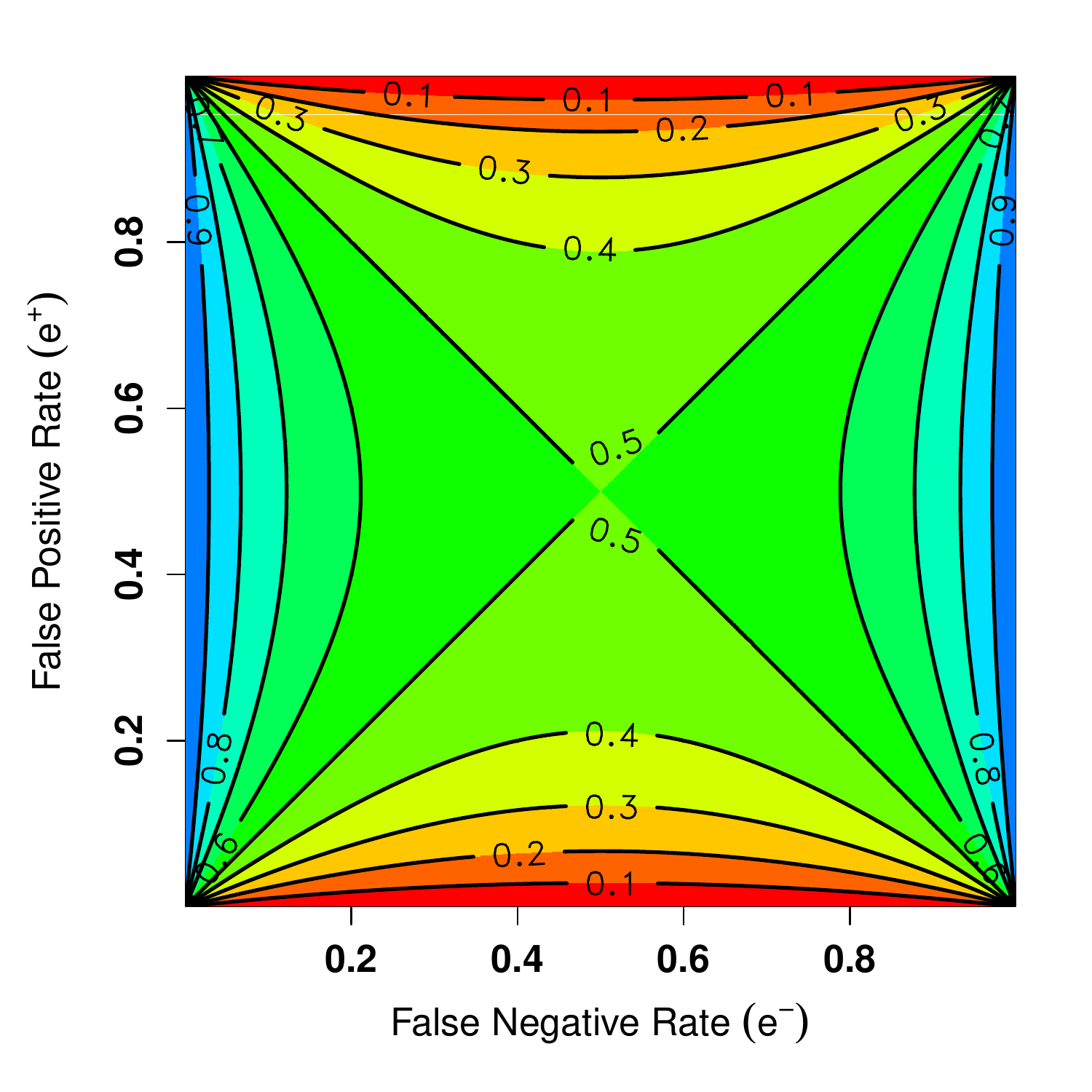}
\caption{Maximum density at which $\ILAS$ has lower expected Hamming error than $\ULAS$, as a function of error rate.  For sufficiently sparse graphs, \ILAS can be superior even when false negative rates are high.  (Note this pattern reverses when $\emp+\epp>1$, in which case individuals are perversely informative.) \label{f_ilas_critden}}
\end{center}
\end{figure}

There is, however, another aspect to this problem.  Consider the expected total (Hamming) errors for \ILAS and \ULAS given the true graph state:
\begin{gather}
\mathbf{E} \sum_{(i,j)\in \mathcal{D}} \left|\ILAS-\Theta\right| = \sum_{(i,j)\in \mathcal{D}} \left[\Theta_{ij} \left(\emp_i+\emp_j-\emp_i\emp_j\right) + \left(1-\Theta_{ij}\right)\epp_i\epp_j\right] \label{e_errilas}\\
\mathbf{E} \sum_{(i,j)\in \mathcal{D}} \left|\ULAS-\Theta\right| = \sum_{(i,j)\in \mathcal{D}} \left[\Theta_{ij} \emp_i\emp_j + \left(1-\Theta_{ij}\right)\left(\epp_i+\epp_j-\epp_i\epp_j\right)\right] \label{e_errulas}
\end{gather}
where $\mathcal{D}$ is the set of potential edges. As a simplifying assumption, let us take all informants to have the same error rates (hence $\emp_i=\emp_j=\emp$ and $\epp_i=\epp_j=\epp$ for all $i,j$).  Let $M(\Theta)=\sum_{(i,j)\in \mathcal{D}} \Theta_{ij}$ be the number of edges in the true graph, and $N(\Theta)=\left|\mathcal{D}\right|-M(\Theta)$ the corresponding number of nulls.  In this scenario, Eq.~\ref{e_errilas} and \ref{e_errulas} reduce to 
\begin{gather*}
\mathbf{E} \sum_{(i,j)\in \mathcal{D}} \left|\ILAS-\Theta\right| = M(\Theta) \left(2\emp-\emp^2\right) + N(\Theta)\epp^2 \\
\mathbf{E} \sum_{(i,j)\in \mathcal{D}} \left|\ULAS-\Theta\right| = M(\Theta) \emp^2 + N(\Theta)\left(2\epp-\epp^2\right).
\end{gather*}
Where the numbers of edges and nulls are similar (i.e., near density 0.5), the edgewise logic invoked above holds.  But this is not always the case.  To see why, let us consider the case in which the expected Hamming error under the intersection rule is greater than under the union rule, i.e.
\begin{align}
\left[\mathbf{E} \sum_{(i,j)\in \mathcal{D}} \left|\ILAS-\Theta\right|\right] &> \left[\mathbf{E} \sum_{(i,j)\in \mathcal{D}} \left|\ULAS-\Theta\right|\right]. \nonumber\\
\intertext{From the above, it follows that this condition corresponds to}
M(\Theta) \left(2\emp-\emp^2\right) + N(\Theta)\epp^2 &> M(\Theta) \emp^2 + N(\Theta)\left(2\epp-\epp^2\right), \nonumber \\
\intertext{and hence}
M(\Theta) \left[\emp-\emp^2 \right] &> N(\Theta) \left[\epp-\epp^2 \right] \nonumber \\
\frac{M(\Theta)}{N(\Theta)} &> \frac{\epp(1-\epp)}{\emp(1-\emp)}, \label{e_ilascond}
\end{align}
which implies that, for sparse graphs ($N(\Theta)\gg M(\Theta)$) the intersection LAS may outperform the union LAS even when $\emp>\epp$.  Intuitively, this is is because the number of \emph{false positive opportunities} is much larger than the number of false negative opportunities in a sparse graph---since the intersection LAS suppresses false positives, it eventually becomes favored at sufficiently low densities.  This is illustrated graphically in Figure~\ref{f_ilas_critden}, which shows the maximum density at which \ILAS outperforms \ULAS under the simplified scenario of Eq.~\ref{e_ilascond}.  As expected from the edgewise analysis, the impact of relative error rates is dominant on LAS performance for moderate densities; however, for densities below $\approx 0.1$, the intersection LAS outperforms the union LAS over a very wide range of informant error rates.  Since densities this low (or lower) are extremely common in social network studies, this suggests that the intersection LAS may outperform the union LAS in many real-world settings.


\section{An Empirical Study of LAS Accuracy}

Although the above theoretical results are intuitive, they depend upon a number of simplifying assumptions.  How does LAS accuracy play out in the real world?  To address this, we conduct an empirical analysis of cognitive social structure data from several studies of two types of relations (friendship and advice-seeking ties) in organizations.  We use statistical models based on the complete data (i.e., all informant reports on all edge variables) to estimate the underlying network structure, and then assess the relative accuracy of \ILAS and \ULAS based on these estimates.  As we show, the empirical results from this analysis are consonant with our theoretical results regarding the superiority of the intersection LAS in sparse graphs.

\subsection{Data and Methods}

To estimate the true network, we utilize a variant of the Bayesian Network Accuracy Model (BNAM) from \citet{butts:sn:2003} with the efficiency-enhancing graph mixture priors of \citet{butts:tr:2017}, where we separately estimate the error rates of each informant self versus proxy ties. The foundations of the model are similar to that of cultural consensus theory \citep{romney.et.al:aa:1986}, and more specifically the generalized Condorcet model described in \citet{batchelder.romney:p:1988}.  As in these models, each informant is assumed to report on the state of each variable (here, each potential $i,j$ edge), erring independently with rates that depend upon the informant and the true state.  In the BBNAM case, these are parameterized as informant-specific false positive and false negative rates (as described in section~\ref{sec_theory}).  Priors are placed on the error rates, and on the underlying graph structure; the former are here taken to be iid beta distributions, and the latter a continuous mixture of $U|man$ graphs of varying density and reciprocity.  The resulting joint posterior provides estimates of both informant error rates and the true graph, given $Y$ (the array of informant reports). Formally, the model can be described as
\begin{gather*}
p(\gamma|\alpha_\gamma) = \mathrm{Dirichlet}(\gamma|\alpha_\gamma)\\
\Pr(\Theta|\gamma) = \prod_{{j,k} \in \mathcal{D}'} \left[\Theta_{jk}\Theta_{kj}\gamma_1 + \left(\Theta_{jk}(1-\Theta_{kj})+(1-\Theta_{jk})\Theta_{kj}\right)\gamma_2 + (1-\Theta_{jk})(1-\Theta_{kj})\gamma_3\right] \\
\begin{split}
\Pr(Y|\epp,\emp,\Theta) =& \prod_{i=1}^N \prod_{(j,k)\in \mathcal{D}} \left[\Theta_{jk} \left[Y_{ijk} (1-\emp_{ijk}) + (1-Y_{ijk})\emp_{ijk}\right] \right.\\
&+ \left. \left(1-\Theta_{jk}\right)\left[Y_{ijk} \epp_{ijk} + (1-Y_{ijk})(1-\epp_{ijk})\right]\right]
\end{split}\\
\end{gather*}
where $\alpha_\gamma$ is a 3-vector of hyperparameters, $\gamma$ is a fully latent vector of prior dyad type frequencies, $\mathcal{D}'$ is the set of undirected dyads (in contrast to the set of directed dyads, $\mathcal{D}$), and other terms are as defined above.  The assumption of homogenity of (respectively) self versus proxy report error rates gives us
\begin{gather*}
\epp_{ijk} =\begin{cases} \epp_{is} & \mathrm{if } i \in \{j,k\}\\ \epp_{ip} & \mathrm{otherwise} \end{cases}\\
\emp_{ijk} =\begin{cases} \emp_{is} & \mathrm{if } i \in \{j,k\}\\ \emp_{ip} & \mathrm{otherwise} \end{cases},
\end{gather*}
where $\epp_{is},\emp_{is}$ are respectively $i$'s self-report error rates, and  $\epp_{ip},\emp_{ip}$ are $i$'s proxy report error rates.  We complete the model by placing independent Beta priors on these rates, giving
\begin{multline*}
p(\epp,\emp|\alpha_{\epp},\beta_{\epp},\alpha_{\emp},\beta_{\emp}) = \prod_{i=1}^n \left[ \mathrm{Beta}\left(\epp_{is}|\alpha_{\epp_s},\beta_{\epp_s}\right) \mathrm{Beta}\left(\emp_{is}|\alpha_{\emp_s},\beta_{\emp_s}\right) \right.\\
\left.\mathrm{Beta}\left(\epp_{ip}|\alpha_{\epp_p},\beta_{\epp_p}\right) \mathrm{Beta}\left(\emp_{ip}|\alpha_{\emp_p},\beta_{\emp_p}\right) \right].
\end{multline*}
Samples from the joint posterior are obtained by a Metropolis within Gibbs algorithm, following the strategy of \citet{butts:tr:2017} as described below.

To compare LAS estimators, we employ data collected from eight networks in four studies: one of an entrepreneurial firm, two of technological firms, and one of a university group, hereby referred to respectively as ``Silicon Systems,'' ``High Tech Managers,'' ``Italian University,'' and ``Pacific Distributors'' \citep{krackhardt:sn:1987, krackhardt:ch:1992, casciaro:1998, kilduff:2008}. Each of the four studies contains CSS data on advice-seeking and friendship relations. Silicon Systems (SS) was originally analyzed with 36 informants, but information on informants 13, 24, and 35 was missing for both friendship and advice networks, leaving a total of 33 informants.\footnote{The three non-responding informants and others' perceptions of their ties were removed upon the release of this dataset to the public, and thus inference cannot be done to approximate the data we would have seen. We treat them as absent by design.} Italian University (IU) contains 25 informants. Pacific Distributors contains 47 informants. As organizations are common contexts for network studies, we regard these settings as broadly comparable to others commonly used by network researchers. We model both friendship and advice-seeking as directed relations, since both were collected using directed prompts (e.g. ``to whom does X go for help and advice at work,'' ``whom does Y consider a friend'').

Informant self-report and proxy report error rates for all eight networks are modeled as drawn a priori from iid Beta distributions as described above, with $\alpha$ and $\beta$ hyperparameters set to 1 and 11, respectively.  As noted above, we employ the $U|man$ mixture prior for the network structure, with a Jeffreys hyperprior on the dyad type rates; this provides a minimally informative, ``flat'' prior with respect to density, reciprocity, and fine structure.  The BNAM utilizes a MCMC procedure to jointly estimate the informant error rate parameters as well as the criterion graph.  We take and discard 500 draws for our burn-in sample, and then sample additional 1000 draws from the joint posterior distribution in each case. We verify MCMC convergence via the $\hat{R}$ statistic.

We employ the posterior network draws from the BNAM as our criterion, comparing the LAS intersection and LAS union against each draw; the distribution of error rates over posterior draws is thus a posterior estimate of the respective accuracies of \ILAS and \ULAS on each data set. The full BNAM estimate on the complete set of informant self and proxy reports represents the best available estimate of the total graph  - and has been shown to be highly accurate in simulation studies \citep{butts:sn:2003, butts:tr:2017} - providing a basis for assessing the much more information-limited LAS estimates.  We measure the error of the LAS estimates versus the estimated criterion by the Hamming distance between the two graphs (i.e., the number of edge variables that would have to be changed to convert one into the other).  The LAS estimator with lower posterior expected Hamming error is estimated to be more accurate for the case in question.

\subsection{Results}

We begin our presentation of results by considering our informants' estimated error rates (since these are expected to play a major role in LAS performance); LAS performance itself is then presented in section~\ref{sec_lasperf}.

\subsubsection{Error Rates} \label{sec_error}

\begin{figure}
\begin{center}
\includegraphics[width=4in]{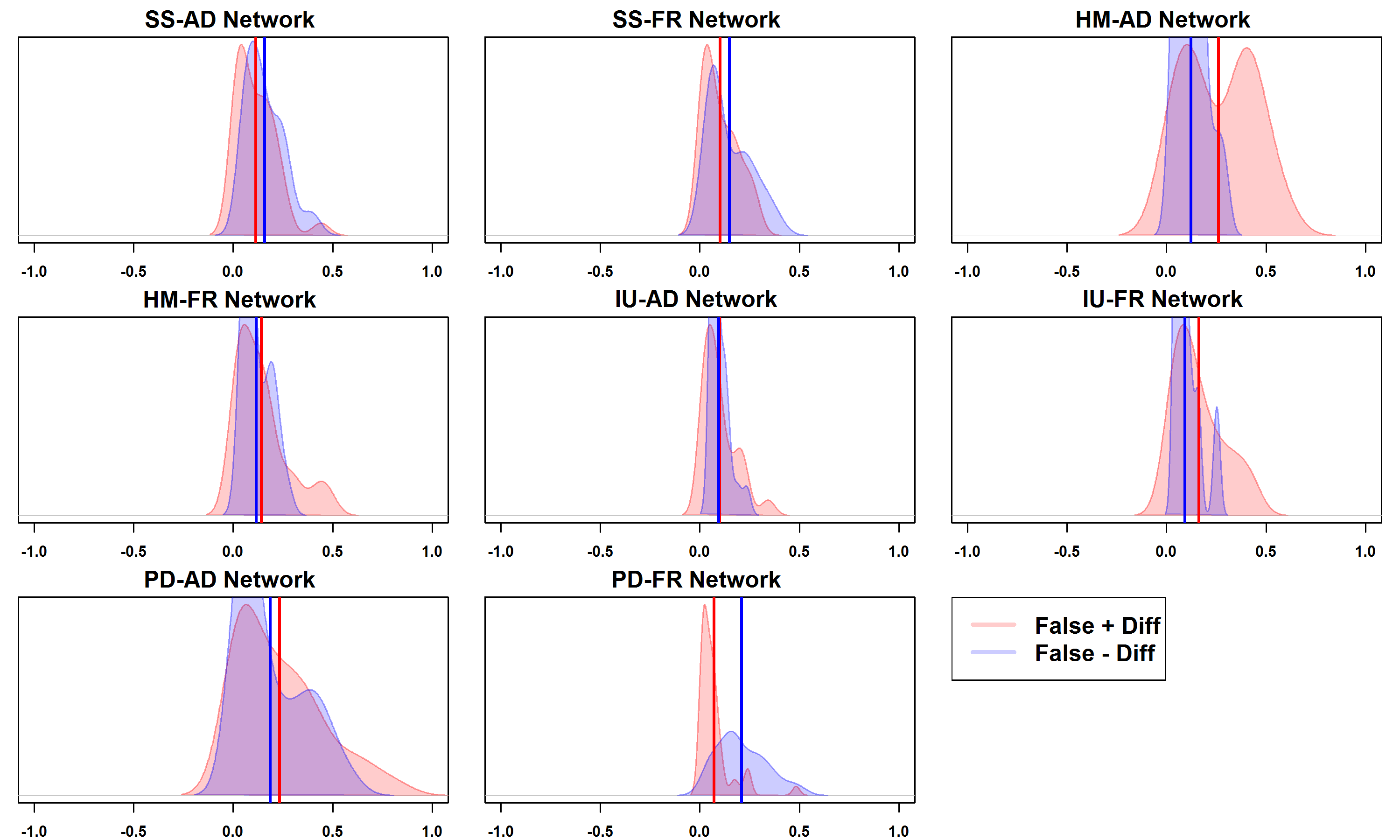}
\caption{Distribution of differences between informants' posterior mean error rates when reporting on their own ties versus reporting on ties between third parties.  More positive numbers indicate higher error rates for self-report versus proxy reports; 0 indicates equal error rates.  Means are shown by vertical lines.  Study labels refer respectively to High-tech Managers (HM), Silicon Systems (SS), Pacific Distributors (PD), and Italian University (IU). The two relations measured were advice-seeking (AD) and friendship (FR).  \label{globaldiff}}
\end{center}
\end{figure}

We begin by taking posterior draws for the BNAM for each network, and examining the informant error rates; as shown in section~\ref{sec_formal}, these are expected to be strongly related to LAS performance.  One immediate observation is that there are substantial differences in false negative rates for proxy vs. self-reports in all networks (Figure~\ref{globaldiff}).  While there is considerable individual variability, individuals are on average substantially more likely to miss ties when reporting on others than when reporting on themselves.  By turns, false positive rates are similar on average for self versus proxy reports, although there is a small but consistent tendency towards over-reporting of self-ties in all networks; the presence of a somewhat elongated right tail in each network also suggests that some individuals are especially prone to over-reporting for ties involving themselves vs. third party ties.  This last result is in line with the conclusions by \citet{kumbasar.et.al:ajs:1994}, who found evidence that informants have a higher propensity to make false positive errors when describing their own ties. 

\begin{figure}
\begin{center}
\includegraphics[width=5in]{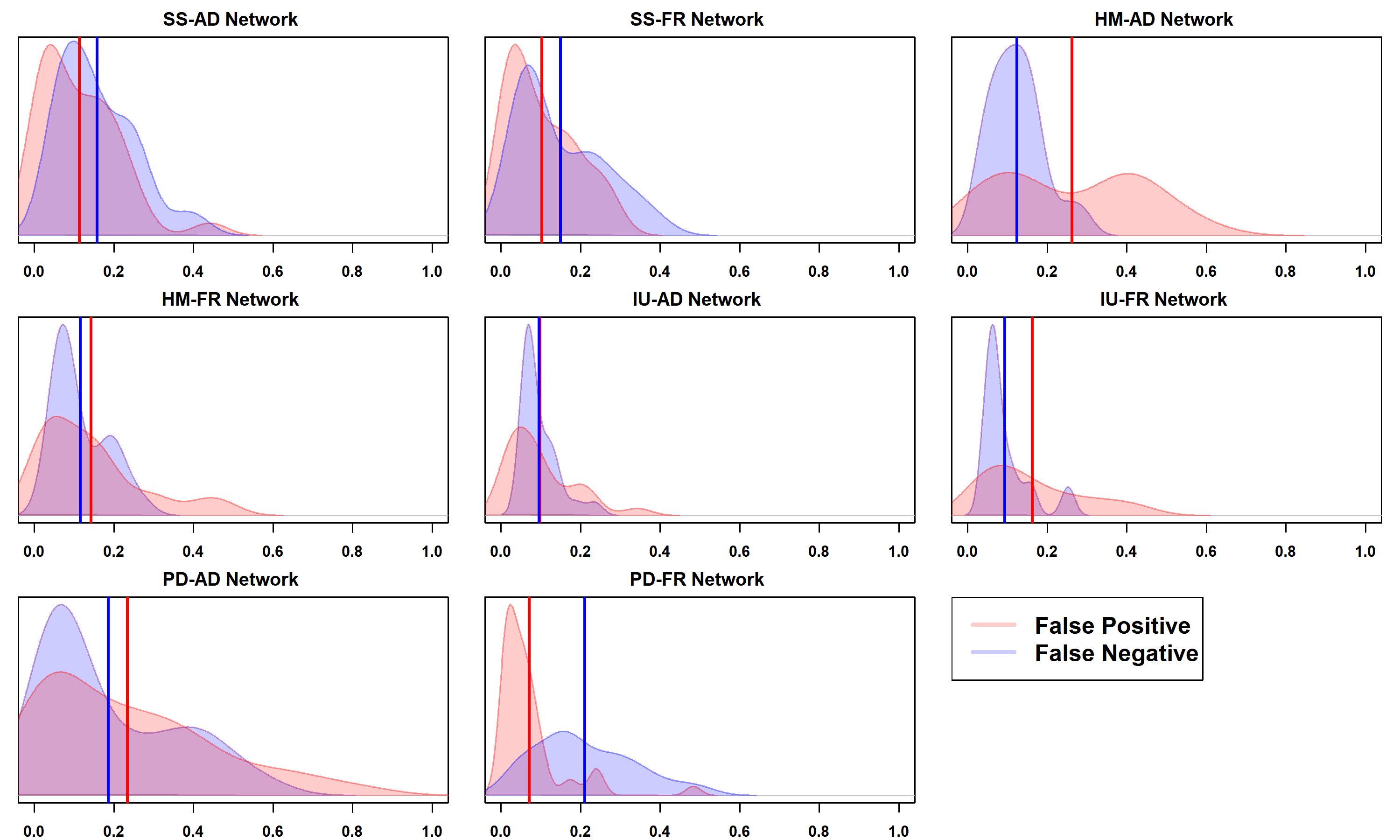}
\caption{ Marginal posterior distributions of informant false positive (FP) and false negative (FN) self-report error rates for each of the 4 organizational settings: Silicon Systems (SS), High-tech Managers (HM), Italian University (IU), and Pacific Distributors (PD). Means indicated by vertical lines.
\label{f_fpfn}}
\end{center}
\end{figure}

\begin{figure}
\begin{center}
\includegraphics[width=5in]{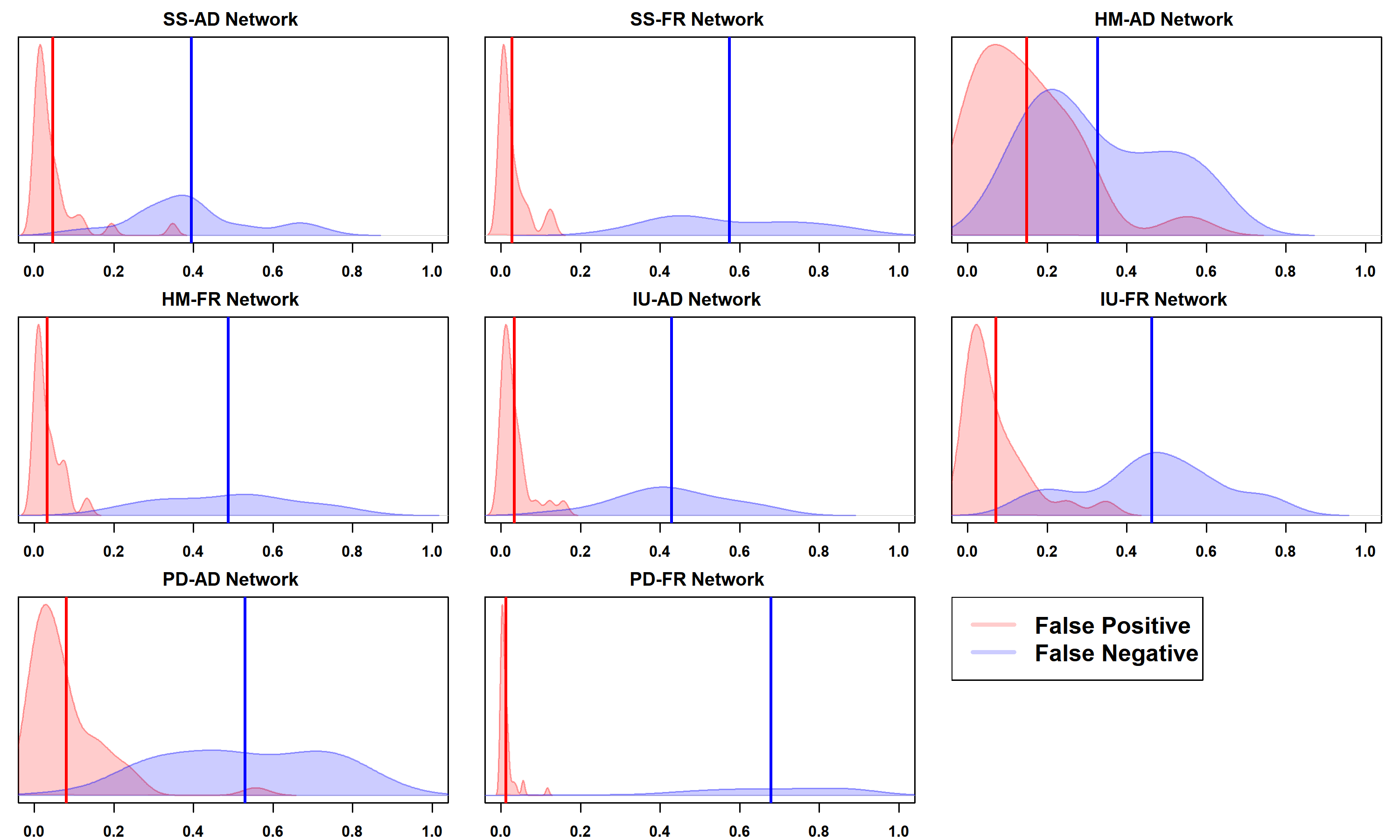}
\caption{ Marginal posterior distributions of informant false positive (FP) and false negative (FN) proxy error rates for each of the 4 organizational settings: Silicon Systems (SS), High-tech Managers (HM), Italian University (IU), and Pacific Distributors (PD). Means indicated by vertical lines.
\label{f_fpfn_proxy}}
\end{center}
\end{figure}

Figure~\ref{globaldiff} shows the differences between self and proxy reporting, but not absolute error levels; these are presented in Figure~\ref{f_fpfn}.   As LAS performance depends only on self-report error rates, we focus on those here. Overall, we see that the two error rates seem to be similar in most networks, with false negatives being slightly more prevalent on average.  In one case (Pacific Distributors friendship) we see much higher false negative rates than false positive rates for most informants (a pattern typical of proxy report error rates), but the reverse is found in two cases (High-tech Manager advice and Italian University friendship).  It is noteworthy that these outlying cases are associated neither with organizational context nor relation, suggesting that they stem from idiosyncratic factors.  It is useful to contrast this situation with that for proxy reports, which are shown for the same networks in Figure~\ref{f_fpfn_proxy}. As one can observe in the proxy error reports, the false negative error rates dwarf the false positive error rates in magnitude. This is consistent with the general trend in global error rates estimated in \citet{butts:sn:2003}. As most reports in a CSS datum will be proxy reports, it stands to reason that the global error rates derived from \citet{butts:sn:2003} would be closer to the proxy error rates that we display here.

To the extent that false positive and false negative error rates are similar for self-report, our theoretical results suggest that the best-performing LAS will be governed by sparsity.  To examine this question, we now turn to our empirical performance comparison.

\subsubsection{LAS Performance Comparison} \label{sec_lasperf}
\begin{figure}
\begin{center}
\includegraphics[width=5in]{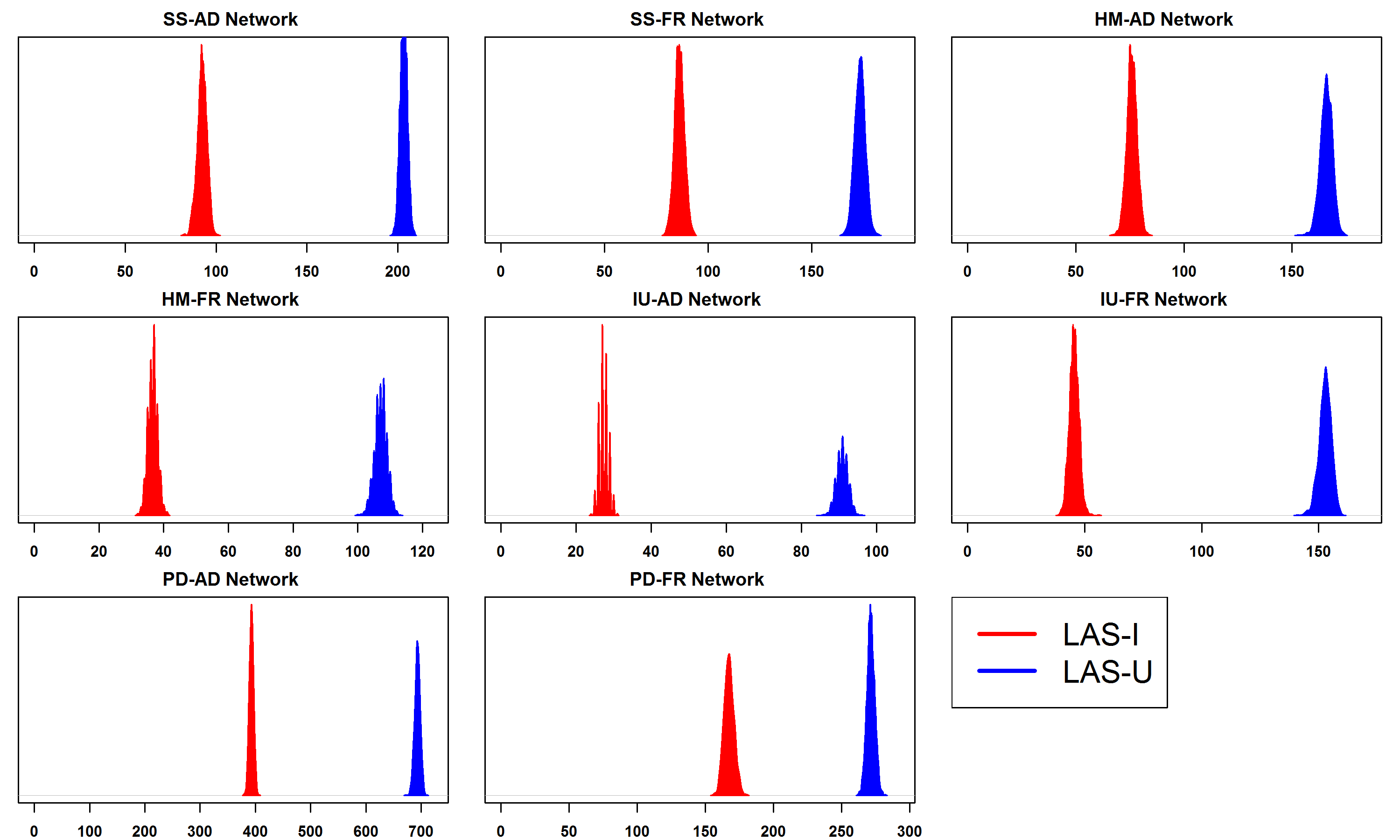}
\caption{Posterior Hamming error distribution for the LAS Intersection and LAS Union methods relative to the criterion graph estimate (from the BNAM) for the various datasets, separated by environment and type of relation measured. The four environments were High-tech Managers (HM), Silicon Systems (SS), Pacific Distributors (PD), and Italian University (IU). The two relations measured were advice-seeking (AD) and friendship (FR). \label{LASPerformance}}
\end{center}
\end{figure}

We present the Hamming error distribution of the two aggregation methods relative to the criterion in Figure~\ref{LASPerformance},s with the mean Hamming error of the two methods being presented in Table~\ref{t_hamerr_sr}. Although we focus on the performance of the LAS methods for aggregation, for completeness, we present the Hamming error distribution of the networks estimated from naive self-report (i.e., own report of outgoing ties) in Figure~\ref{sec_lasperf}.
We see that the LAS intersection consistently provides a more accurate estimate of the criterion graph relative to the LAS union. In addition, the posterior Hamming error distributions are fairly well-localized around their means. 

\begin{table}
\begin{center}
\begin{tabular}{crrr} \hline\hline
Dataset & Union & Intersection & Self-Report\\ \hline
HM-AD & 166 & 75 & 124\\
HM-FR & 106 & 35 & 72\\
SS-AD & 204 & 93 & 141\\
SS-FR & 173 & 87 & 121\\
PD-AD & 693 & 392 & 583\\
PD-FR & 271 & 167 & 232\\
IU-AD & 91 & 27 & 58\\
IU-FR & 155 & 46 & 101\\ \hline\hline
\end{tabular}
\caption{Hamming error of the LAS Union or LAS Intersection aggregation relative to the criterion graph (estimated as the BNAM central graph) for the various datasets, separated by environment and type of relation measured. The four environments were High-tech Managers (HM), Silicon Systems (SS), Pacific Distributors (PD), and Italian University (IU). Lower values indicate better performance.  The two relations measured were advice-seeking (AD) and friendship (FR). \label{t_hamerr_sr}}
\end{center}
\end{table}



The definitive superiority of the LAS Intersection rule despite the relative similarity of self-report error rates suggests an origin in sparsity (i.e., having more opportunities to commit false positive vs. false negative errors).  As Table~\ref{t_density} shows, our networks are indeed fairly sparse, with estimated densities less than 0.3 in all cases.  Thus, even in the Pacific Distributor friendship network, where the average false negative rate was more than twice as large as the average false positive rate, we see that there were nearly nine times as many opportunities for these false positives to occur as there were for false negatives.  One hence obtains superior performance from suppressing the error that informants \emph{have more chances to make,} as seen in Figure~\ref{LASPerformance}.  Although extremely high ratios of false negative to false positive rates would lead to the opposite result, Table~\ref{t_density} shows that these rate differences would need to be very large to overcome the opportunity effect.

\begin{table}
\begin{center}
\begin{tabular}{ccc} \hline\hline
Dataset & Density & FP/FN Opportunities\\ \hline
HM-AD & 0.286 & 2.50\\
HM-FR & 0.119 & 7.40\\
SS-AD & 0.123 & 7.13\\
SS-FR & 0.121 & 7.26\\
IU-AD & 0.096 & 9.42\\
IU-FR & 0.157 & 5.37\\
PD-AD & 0.257 & 2.89\\
PD-FR & 0.112 & 7.93\\ \hline\hline
\end{tabular}
\caption{Posterior mean density estimates of the criterion network via the BNAM, separated by environment and type of relation measured. The four environments were High-tech Managers (HM), Silicon Systems (SS), Pacific Distributors (PD), and Italian University (IU). The two relations measured were advice-seeking (AD) and friendship (FR). All networks are sparse, providing from 2.5 to almost 9.5 times as many opportunities for false positive errors than false negative errors (third column). \label{t_density}}
\end{center}
\end{table}

\section{Discussion and Conclusion}

With this study we sought to determine which LAS method would reproduce an unknown criterion graph more accurately. Across all eight of the networks examined here, we found that the LAS Intersection outperformed the LAS Union. A cursory examination of the density of the graphs indicate that they are all relatively sparse. As the sparsity of the analyzed graphs serves to create more opportunities for false positives, this creates conditions that are more favorable for the LAS Intersection. Although this would be outweighed by a sufficiently high false negative rate (relative to the false positive rate), this disparity must become quite large to overcome the levels of sparsity observed in typical settings.  Importantly, we do not find evidence of this disparity in self-report errors, implying that the LAS Union is unlikely to be superior except in networks for which the density approaches or exceeds 50\%.  Our findings are in line with Equation~\ref{e_ilascond}, which gives an approximate criterion for the conditions under which one or the other rule is expected to dominate.

It should be noted that the results given here focus exclusively on Hamming error---i.e., the number of edge variables incorrectly inferred---and may not apply to all analyses.  In particular, network measures that are differentially sensitive to false positive vs. false negative errors may have different optimal procedures.  When in doubt, a simulation study should be performed prior to analysis.

Although one LAS may be better than another in a given setting, this does not imply that the LAS is per se the best tool to use: indeed, as shown in Figure~\ref{LASPerformance}, even the Intersection LAS shows high levels of Hamming error in all cases examined here.  Where possible, we recommend obtaining as many measures of each edge variable as possible (either via complete CSS \cite{krackhardt:sn:1987} or arc sampling designs \cite{butts:sn:2003}) and subjecting them to model-based inference.  In many cases, however, analysts must make due with only self-report data.  In these settings, the Intersection LAS will often outperform the Union LAS, and we recommend the former over the latter so long as the networks in question are suitably sparse.



\bibliography{bibhome/las_compare} 


\end{document}